\documentstyle[aps,preprint]{revtex}

\begin{document}
\draft
\preprint{INR-0969/98, PURD-TH-98-04}
\date{February 1998}
\title{Ultra-High Energy Cosmic Rays, Superheavy Long-Living Particles,
and Matter Creation after Inflation}
\author{Vadim Kuzmin$^{a,}$\footnote{email address: 
kuzmin@ms2.inr.ac.ru}  and Igor Tkachev$^{a,b,}$\footnote{ email address:
tkachev@physics.purdue.edu }}
\address{
$^a$Institute for Nuclear Research, Russian Academy of Sciences,\\ 
60th October Anniversary Prosp. 7a, Moscow 117312, RUSSIA \\
$^b$Department of Physics, Purdue University, West Lafayette, IN 47907, USA
}

\maketitle
 
\begin{abstract}
The highest energy cosmic rays, above the  Greisen--Zatsepin--Kuzmin cut-off 
of cosmic ray spectrum,  may be produced in decays of superheavy 
long-living $X$-particles. We conjecture that these particles may be 
produced {\it naturally} in the early Universe from {\it vacuum fluctuations} 
during inflation  and may  constitute a considerable fraction of Cold Dark 
Matter.
We predict a new cut-off in the UHE cosmic ray spectrum
$E_{\rm cut-off} < m_{\rm inflaton} \approx 10^{13}$ GeV, the exact 
position of the cut-off and the shape of the cosmic ray spectrum beyond the 
GZK cut-off being determined by the QCD quark/gluon fragmentation. The Pierre
Auger Project installation  might discover this phenomenon.
\end{abstract}


\narrowtext

According to the Greisen-Zatsepin-Kuzmin \cite{gzk} (GZK) bound, the
Ultra High Energy (UHE) cosmic rays produced in any known candidate 
extra galactic source should have an
exponential cut-off at energies $E \sim 5 \times 10^{10}$ GeV.
On the other hand, the number of observed \cite{cr} cosmic rays events beyond
the cut-off is growing and leads to a mounting paradox within standard 
frameworks of cosmological and particle physics models.

A wide variety of possible solutions were  suggested. Resolution could be due 
to exotic particle which may be produced
at cosmological distances were suitable conventional accelerators
are found, be transmitted evading GZK bound, and yet which interact in the 
atmosphere like a hadron. A particle with correct properties was found
in a class of supersymmetric theories \cite{farrar}. Alternatively, 
high energy cosmic rays may have been produced locally. One possibility
is connected to the events of destruction of (topological) defects
\cite{td}, while another one to decays of primordial heavy long-living
particles \cite{KR,BKV}. The candidate particle must obviously  obey 
constraints on the mass, density and lifetime. 

In order to produce cosmic rays of energies $E \agt 10^{11}$ GeV, the mass of 
$X$-particles has to be very large, $m_{X} \agt 10^{13}$~GeV \cite{KR,BKV}.
The lifetime, $\tau_{X}$, cannot be much smaller than 
the age of the Universe, $\tau \agt 10^{10}$~yr. With this smallest
value of the lifetime, the observed flux of UHE cosmic rays will be
reproduced with rather low density of $X$-particles, 
$\Omega_{X} \sim 10^{-12}$,
where $\Omega_{X} \equiv m_{X} n_{X}/\rho_{\rm crit}$, $n_X$ is the 
number density
of X-particles and $\rho_{\rm crit}$ is the critical density.  
On the other hand, X-particles must not overclose the Universe, 
$\Omega_{X} \alt 1$.
With $\Omega_{X} \sim 1$, the X-particles may play the role of cold dark 
matter and the observed flux of UHE 
cosmic rays can be matched if $\tau_{X} \sim 10^{22}$~yr. 
The allowed windows are quite wide \cite{KR}, but on exotic side, which may
rise problems.  

The problem of the particle physics mechanism responsible for a long but 
finite lifetime of very heavy particles can be solved in several ways.
For example, otherwise conserved quantum number carried by the X-particle 
may be broken very weakly 
due to instanton transitions \cite{KR}, or quantum gravity (wormhole) 
effects \cite{BKV}. If instantons are responsible for $X$-particle 
decays, the lifetime is roughly estimated as 
$\tau_{X} \sim m_{X}^{-1}\cdot \mbox{exp}(4\pi/{\alpha_{X}})$,
where $\alpha_{X}$ is the coupling constant of the relevant (spontaneously 
broken) gauge symmetry. Lifetime will fit the window if 
the coupling constant (at the scale $m_{X}$) is 
$\alpha_{X} \approx 0.1$ \cite{KR}.

X-particles can be produced in the right amount by usual collision and
decay processes if the reheating temperature after inflation never exceeded 
$m_X$, but the temperature should be in the range 
$10^{11} \alt T_r \alt 10^{15}$ GeV, depending on $m_{X}$, \cite{KR,BKV}.
This is a rather high value of reheating temperature and will lead to the
gravitino problem in generic supersymmetric models \cite{gtino}.

In the present paper we investigate another process of X-particle creation,
namely the direct production from vacuum fluctuations during inflation.

Any viable modern cosmological model invokes the hypothesis of 
inflation \cite{inflation}. During inflation the Universe expands 
exponentially, 
which solves the horizon and flatness problems of the standard Big-Bang 
cosmology. Inflation is generally assumed to be driven by the special 
scalar field $\phi$ known as the {\it inflaton}. Fluctuations generated at 
inflationary stage can have strength and the power spectrum suitable for
generation of the large scale structure. This fixes the range of 
parameters in the inflaton potential. For example, the mass of the inflaton 
field has to be $m_\phi \sim 10^{13}$ GeV. During inflation, 
the inflaton field 
slowly rolls down towards the minimum of its potential. Inflation ends when 
the potential energy associated with the inflaton field became smaller than 
the kinetic energy.  At that time all the energy of the Universe is contained 
entirely 
in the form of coherent oscillations of the inflaton field around the minimum 
of its potential. It is possible that a significant fraction of this energy 
is released to other Boson species after only a dozen or so inflaton field 
oscillations, in the regime of a broad parametric resonance \cite{KLS}.
This process was studied in details \cite{KT,KLS2}. Even particles with 
masses of order
of magnitude larger than the  inflaton mass can be produced quite abundantly.
Applying these results to the case of our interest here, we find
that stable very heavy particles, $m_X \agt m_\phi$, generally will 
be produced in excess and will overclose the Universe.

However, if the parametric resonance is ineffective for some reason, and we
estimate particle number density after inflation at the level of initial 
conditions used in Refs. \cite{KT}, we find that $\Omega_X$ 
might prove to be of about the right magnitude. This level is saturated 
by the fundamental process of particle creation during inflation from 
{\it vacuum fluctuations} and it is the same process which generated 
primordial large scale density fluctuations.
Parametric resonance for $X$ particles is turned-off if $X$ is either
a fermion field or its coupling to inflaton is small,
$g^2 \ll 10^4 (m_X/m_\phi)^4 (m_\phi/M_{\rm Pl})^2 $ \cite{KT}.

At some early epoch the metric of the Universe is conformally flat to a high
accuracy, $ds^2 = a(\eta )^2 (d\eta^2 - d{\bf x}^2)$. We normalize the scale 
factor by the condition $a(0)=1$ at the end of inflation.  
Number density of particles created in time varying cosmological 
background can be written as
\begin{equation}
n_X=\frac{1}{2\pi^2a^3} \int |\beta_k|^2 k^2 dk \, \, ,
\label{nX}
\end{equation}
where $\beta_k$ are the Bogoliubov coefficients which
relate ``in'' and ``out'' mode functions, and $k$ is the comoving momentum. 
Massles conformally coupled particles (for scalars this means that
$\xi =1/6$ in the direct coupling to the curvature) are not 
created. For massive particles conformal invariance is broken. Therefore, for
the power low (e.g., matter or radiation dominated) period of expansion of 
the Universe, one expects on dimensional grounds, $n_X \propto m_X^3/a^3$
at late times.
Indeed, it was found in Ref. \cite{pc_pwl}
\begin{equation}
n_X \approx 5.3 \times 10^{-4} m_X^3 \, (m_X t)^{-3/2} \, \, ,
\label{nXrd}
\end{equation}
for the radiation dominated Universe, and $n_X \propto m_X^3 \, (m_X t)^{-3q}$
for $a(t) \propto t^q$. Note that all particle creation occur
in the region $mt  = qm/H \alt 1$. When $mt \ll 1$, the number density of 
created 
particles remains on the constant level $n_X = m_X^3/24\pi^2$ independent of
$q$ \cite{pc_pwl}. At $qm/H \gg 1$ particle creation is negligible.
Here H is the Hubble constant, $H \equiv \dot{a}/a$.

For the radiation dominated Universe one finds, $\Omega_X \sim  
(m_X^2/M^2_{\rm Pl})\sqrt{m_X t_e}$, 
where $t_e$ is time of equal densities of radiation 
and matter in $\Omega =1$ Universe. This gives $\Omega_X \sim m_9^{5/2}$,
where $m_9 \equiv m_X/10^9$ GeV. Stable particles with $m_X \agt 10^9$ GeV
will overclose the Universe even if they were created from the vacuum during
regular Friedmann radiation dominated stage of the evolution.
It is possible to separate vacuum creation from creation in collisions 
in plasma since X-particles may be effectively sterile.

This restriction can be overcomed if evolution of the Universe,
as it is believed, was more complicated than simple radiation dominated 
expansion from singularity. Hubble constant may have
never exceeded $m_X$, which is the case of inflation, $H(0) \approx m_\phi$.
Moreover, compared to the case considered above, 
density of X-particles created during inflation is additionally diluted 
by late entropy release in reheating after inflation.

Particle creation from vacuum fluctuations during inflation (or in de Sitter
space) was extensively studied \cite{pc_dS1,pc_dS2}.
Characteristic quantity which is usually cited, the variance of 
the field $\langle X^2 \rangle$, is defined by an expression similar to 
Eq. (\ref{nX}). In the typical case $\alpha_k \approx -\beta_k$ the 
difference is given by the 
factor $2\sin^2(\omega_k\eta)/\omega_k$ in the integrand, where
$\omega_k^2 = k^2 + a^2m^2_X$.
If $m_X \sim H(0) \approx m_\phi$,
one  has on dimensional grounds $n_X = C{m_\phi^3}/{2\pi^2a^3}$ where the
coefficient $C$ is expected to be somewhat smaller than unity. 
Both Fermions and Bosons are prodused by this mechanism, exact numerical
value of $C$ being dependent on spin-statistics. In general,
$C$ is the function of the ratio $H(0)/m_X$, the function of self-coupling
of $X$ and the coupling $\xi$, depends on details of the transition 
between inflationary and matter (or radiation) dominated phases, etc. 
For example, for the scalar Bose-field, 
$\langle X^2 \rangle = 3H(0)^4/8\pi^2m_X^2$ if 
$m_X \ll H(0)$\cite{pc_dS1,pc_dS2}.  For massless self-interacting field
$\langle X^2 \rangle \approx 0.132 H(0)^2/\sqrt{\lambda}$ \cite{SY}.
$C$ is expected to decrease exponentially when $m_X >m_\phi $.
Particle creation in the case of Hubble dependent effective
mass, $m_X(t) \propto H(t)$, was considered in Ref. \cite{LR}.

Let us estimate the today's number density of X-particles. We consider 
massive inflaton, $V(\phi) = m_\phi^2 \phi^2/2 $. In this case inflation 
is followed by the matter domination stage. If there are light Bosons
in a theory, $m_B \ll m_\phi$, even relatively weakly coupled to the inflaton,
$g^2 \agt 10^4 m_\phi^2/M_{\rm Pl}^2 \sim 10^{-8}$, this matter domination
stage will not last long: inflaton will decay via parametric resonance
and the radiation domination follows. This happens typically when the energy 
density in inflaton oscillations is redshifted by a factor 
$r \approx 10^{-6}$ compared 
to a value $m_\phi^2 M_{\rm Pl}^2$ \cite{KLS,KT}. Matter is still far
from being in the thermal equilibrium, but it is still convenient 
to characterize
this radiation dominated stage by an equivalent temperature, 
$T_* \sim r^{1/4} \sqrt{m_\phi M_{\rm Pl}}$. At this moment the ratio of 
energy density
in X-particles to the total energy density retains its value reached at the 
end of inflation, $\rho_X/\rho_R \approx C\, m_\phi m_X/2\pi^2 M_{\rm Pl}^2$.
Later on this ratio grows as $\propto T/T_*$ and reaches unity
at $T = T_{\rm eq}$, where
\begin{equation}
T_{\rm eq} = \frac{Cr^{1/4}}{2\pi^2} 
\left(\frac{m_\phi}{M_{\rm Pl}}\right)^{3/2}m_X \, .
\label{Teq}
\end{equation}
Using relation $T_{\rm eq} = 5.6 \Omega_X h^2 $ eV we find that
$10^{-12} \alt \Omega_X \alt 1$ if 
\begin{equation}
10^{-23} \alt Cr^{1/4} m_X/m_\phi \alt 10^{-11} \, .
\label{range}
\end{equation}

For $m_X \sim (\rm a~few) \cdot m_\phi$ this condition can be easily 
satisfied since
the coefficient $C$ is exponentially small. This condition may be satisfied
even for  $m_X \sim m_\phi$ since the coefficient $r^{1/4}$
(or equivalent reheating temperature) can be small too. 

Our hypothesis has unique observational consequences. If UHE cosmic rays
are indeed due to decay of superheavy particles which were produced from vacuum
fluctuations during inflation, there has to be a new sharp cut-off in
the cosmic ray spectrum at energy somewhat smaller $m_X$. Since the number
density $n_X$ depends exponentially upon $m_X/m_\phi$, the position of this
cut-off might be well predicted and has to be near
$ E_{\rm cut-off} < m_\phi \approx 10^{13}$ GeV, the very shape of the cosmic
ray spectrum beyond the GZK cut-off being of quite generic form
following from the QCD quark/gluon fragmentation. The Pierre Auger Project
installation \cite{auger}  might prove to be able to discover this
fundamental phenomenon.

We conclude, observation of Ultra High Energy cosmic rays can probe
the spectrum of elementary particles in its superheavy range and can be
an additional opportunity (alongside with fluctuations in cosmic microwave
background) to study the earliest epoch of the Universe evolution, starting 
from amplification of vacuum fluctuations during inflation
through fine details of gravitational interaction and down to physics
of reheating.

When our paper was at the very end of completion we became aware of the 
quite recent paper by Chung, Kolb and Riotto {\cite{CKR}} where similar 
problems of superheavy dark matter creation were considered.

We are grateful to S. Khlebnikov for useful discussions.
V.~A.~Kuzmin and I.~I.~Tkachev thank Theory Division at CERN for 
hospitality where the major part of this work was done.  
The work of V.~K. was supported in part by 
Russian Foundation for Basic Research under  Grant 95-02-04911a.
I.~I.~T.  was supported in part by the U.S. Department of Energy
under Grant DE-FG02-91ER40681 (Task B) and by the National Science
Foundation under Grant PHY-9501458.


\begin{thebibliography}{99}

\bibitem{gzk}
K. Greisen, Phys. Rev. Lett. {\bf 16}, 748 (1966); 
G. T. Zatsepin and 
V. A. Kuzmin, Pisma Zh. Eksp. Teor. Fiz. {\bf 4}, 114 (1966).

\bibitem{cr}
N. Hayashida et.al., Phys. Rev. Lett. {\bf 73}, 3491 (1994);
D. J.Bird et.al., Astroph. J. {\bf 424}, 491 (1994); 
{\bf 441}, 144 (1995); 
T. A.Egorov et.al., in: {\it Proc.  Tokyo Workshop on 
Techniques for the Study
of Extremely High Energy Cosmic Rays}, ed. M.Nagano (ICRR, U. of Tokyo,
1993).

\bibitem{farrar} G. R. Farrar, Phys. Rev. Lett. {\bf 76}, 4111 (1996);
D. J. Chung, G. R. Farrar, and E. W. Kolb, astro-ph/9707036.

\bibitem{td}
C. T. Hill, Nucl. Phys. {\bf B224}, 469 (1983);
C. T. Hill, D. N. Schramm and T. P. Walker, Phys. Rev. {\bf D36}, 
1007 (1987);
G. Sigl, D. N.Schramm and P. Bhattacharjee, Astropart. Phys. {\bf 2}, 
401 (1994);
V. Berezinsky, X. Martin and A. Vilenkin, Phys. Rev. {\bf D56}, 
2024 (1997);
V. Berezinsky and A. Vilenkin, astro-ph/9704257.

\bibitem{KR}
V. A. Kuzmin and V. A. Rubakov, astro-ph/9709187.

\bibitem{BKV} V. Berezinsky, M. Kachelriess and A. Vilenkin,
Phys. Rev. Lett. {\bf 79}, 4302 (1997).

\bibitem{gtino} J. Ellis, J. E. Kim, and D. V. Nanopoulos,
Phys. Lett. {\bf B145}, 181 (1984).

\bibitem{inflation} For a review and list of references, see A. D. Linde,
{\it Particle Physics and Inflationary Cosmology},
(Harwood Academic, New York, 1990); 
E. W. Kolb and M. S. Turner, {\em The Early Universe}, 
(Addison-Wesley, Reading, Ma., 1990).

\bibitem{KLS}
L. A. Kofman, A. D. Linde  and A. A. Starobinsky, Phys. Rev. Lett.
{\bf 73}, 3195 (1994).

\bibitem{KT} S.Yu. Khlebnikov and I.I. Tkachev,
Phys. Rev. Lett. {\bf 77}, 219 (1996); 
Phys. Lett. {\bf B390}, 80 (1997); 
Phys. Rev. Lett. {\bf 79}, 1607 (1997);
Phys. Rev. {\bf D56}, 653 (1997).

\bibitem{KLS2}
L. A. Kofman, A. D. Linde  and A. A. Starobinsky, Phys. Rev. 
{\bf D56}, 3258 (1997).

\bibitem{pc_pwl} S. G. Mamaev, V. M. Mostepanenko, and A. A. Starobinskii,
ZhETF {\bf 70}, 1577 (1976) [Sov. Phys. JETP {\bf 43}, 823 (1976)].

\bibitem{pc_dS1}
N. A. Chernikov and E. A. Tagirov, Ann. Inst. Henri Poinare
{\bf 9A}, 109 (1968);
E. A. Tagirov, Ann. Phys. {\bf 76}, 561 (1973);
T. S. Bunch and P. C. W. Davies, Proc. R. Soc. {\bf A360}, 117 (1978).

\bibitem{pc_dS2}
A. D. Linde, Phys. Lett. {\bf 116B}, 335 (1982);
A. A. Starobinsky, Phys. Lett. {\bf 117B}, 175 (1982);
A. Vilenkin and L. H. Ford, Phys. Rev. {\bf D26}, 1231 (1982);
B. Allen, Phys. Rev. {\bf D32}, 3136 (1985).

\bibitem{SY} A. A. Starobinsky and J. Yokoyama, 
Phys. Rev. {\bf D50}, 6357 (1994).

\bibitem{LR} D. Lyth and D. Roberts, hep-ph/9609441.

\bibitem{auger} M. Boratav, for Pierre Auger Collaboration, 
{\it The Pierre Auger Observatory Project: An Overview.},  
Proc. of 25th International Cosmic Ray Conference, Durban, v. 5, p. 205 (1997).

\bibitem{CKR} D. J. H. Chung, E. W. Kolb and A. Riotto, hep-ph/9802238.


\end{thebibliography}
\end{document}